\documentclass{elsart}

\usepackage{graphicx}

\begin{document}

\begin{frontmatter}

\title{Origin of superconductivity transition broadening in MgB$_2$}
\author[ISTEC,JST]{T. Masui\corauthref{SRL}},
\ead{masui@istec.or.jp}
\author[ISTEC]{S. Lee},
\author[ISTEC]{S. Tajima}
\address[ISTEC]{Superconductivity Research Laboratory, ISTEC, 1-10-13 Shinonome, Tokyo, 135-0062, Japan}
\address[JST]{Domestic Research Fellow, Japan Society for the Promotion Science, Kawaguchi, Japan}
\corauth[SRL]{Corresponding author. Tel.: +81-3-3536-0618; fax: +81-3-3536-5705.}

\date{\today}

\begin{abstract}
We report resistivity and magnetization of single crystal MgB$_2$, 
focusing on the broadening of superconducting (SC) transition 
in magnetic fields. 
In-plane and out-of-plane resistivity indicate that the broadening 
of superconducting transition is independent of Lorentz force 
and that it is merely dependent on the magnetic field direction. 
In magnetization, diamagnetic signal begins to appear at almost 
the same temperature as the onset temperature of resistivity transition. 
These results suggest that the broadening is attributed 
not to the surface superconductivity but 
to the superconducting fluctuation or the vortex-liquid picture, 
owing to the short coherence length and the high transition temperature 
of MgB$_2$. 

\end{abstract}

\end{frontmatter}

\section{Introduction}

Since the discovery of superconductivity in MgB$_2$ \cite{nagamatsu}, 
the upper critical fields have been one of the main interests. 
After succeeding in crystal growth \cite{lee,xu,kim,sologubenko}, 
the anisotropy of critical field has been investigated 
by many research groups \cite{lee,xu,kim}, 
since relatively large anisotropy was expected 
from the layered crystal structure. 
The anisotropy of $H_{c2}$  has been determined through 
the resistivity measurements, followed by specific heat and torque 
magnetometry measurements \cite{Zehetmayer,Welp}. 
Since a superconductivity transition in resistivity broadens 
in magnetic fields, critical temperature ($T_c$) 
has been determined from the onset temperature $T_{on,\rho}$, where 
resistivity starts to drop. 
The obtained anisotropy ratio $H_{c2}^{ab}//H_{c2}^{c}$ of about 3 at 25 K 
is in good agreement among the reports \cite{lee,xu,kim,masui_iss}. 
However, contrary to the resistivity results, 
an ESR study first indicated much larger anisotropy ratio of 6-9 \cite{Simon}. 
One of the origins for this discrepancy in anisotropy ratio is the 
uncertainty of the $T_c$ determination, 
owing to the broadening in magnetic fields. 
Actually, if $T_c$ is determined by the temperature $T_0$ where resistivity 
vanishes, the higher anisotropy ratio of about 5 is obtained.  
This casts a question what is the physical meaning of 
the onset temperature $T_{on}$. 

As to the resistive broadening, many explanations have been proposed; 
glassy state of vortices \cite{HJ_Kim,Pradhan_01,Pradhan_02}, 
surface superconductivity related to $H_{c3}$ \cite{Welp}, and 
two superconductivity gap effect \cite{Pradhan_02}. 
In the report of surface superconductivity \cite{Welp} 
magnetization data played a critical role 
as the evidence for absence of bulk superconductivity, 
where no diamagnetic signal was observed 
in the $T$-range of resistivity broadening. 
Here, however, the small crystal size makes discussion unclear. 
More detailed measurement, especially magnetization study 
on larger single crystals has been desired. 

Broadening of superconductivity transition is observed in various kinds of 
superconductors \cite{ginsberg,ishiguro}. 
It is usually due to a superconducting fluctuation, derived from a short 
coherence length ($\xi$) and a low-dimensionality. 
If we consider the superconducting parameters of MgB$_2$ such as the high 
$T_c \approx$ 40 K and the short coherence length 
$\xi \approx$ 100 \AA \cite{lee,xu} together with the anisotropy, 
it is likely that thermal fluctuation effect manifests itself near $T_c$. 
However, to our knowledge, there has been little report on 
the fluctuation effect in MgB$_2$, 
except for two reports \cite{lascialfari,Park}. 

In this study we present the superconducting transition behavior of 
MgB$_2$ single crystal, via $\rho_{ab}$, $\rho_c$ and magnetization. 
The onset temperature $T_{on}$ 
below which a sign of superconductivity is observed, 
is well determined in both resistivity and magnetization. 
The whole data support that the $H_{c2}$ is strongly affected by a 
thermal fluctuation of superconductivity. 

\section{Experimental}
Single crystals of MgB$_2$ were grown under high-pressure 
in Mg-B-N system \cite{lee}. 
Resistivity was measured by a four-probe method. 
A typical dimension of the samples for in-plane resistivity measurements was 
300$\times$100$\times$30 $\mu$m$^3$.  
The dimension of a crystal for $c$-axis resistivity measurements 
was 200$\times$200 $\mu$m$^2$ in the $ab$-plane and 150 $\mu$m 
along the $c$-axis. 
The temperature dependence of $\rho_c$ above $T_c$ is almost the same 
as that of $\rho_{ab}$, and the estimated anisotropy 
$(\rho_{c}/\rho_{ab})^{0.5}$ is about 3-7 \cite{masui_iss}. 
Electrical current ($I$) was changed from 0.05 mA up to 10 mA. 
A magnetic field up to 12 Tesla (T) was applied both parallel 
and perpendicular to the $c$-axis, 
and resistivity was measured by sweeping the temperature down to 5 K. 
In magnetization measurement, a commercial SQUID magnetometer 
(Quantum Design) was used. 
We have aligned 100 pieces of crystals on a sample holder, 
in order to gain a magnetization signal. 
The total weight of the sample was about 0.7 mg. 
Each crystal consists of a single domain and has a flat $ab$-surface. 
Magnetization measurements were carried out 
in a warming-up $T$- sweep after zero-field cooling and a cooling-down sweep 
in a constant magnetic field up to 7 T in both directions 
parallel and perpendicular to the $c$-axis. 
We confirmed no sweep-rate dependence of the magnetization by measuring 
with a very slow sweep rate of 4 mK/min. 

\section{Results and discussion}
\begin{figure}
\begin{center}
\includegraphics[width=10cm]{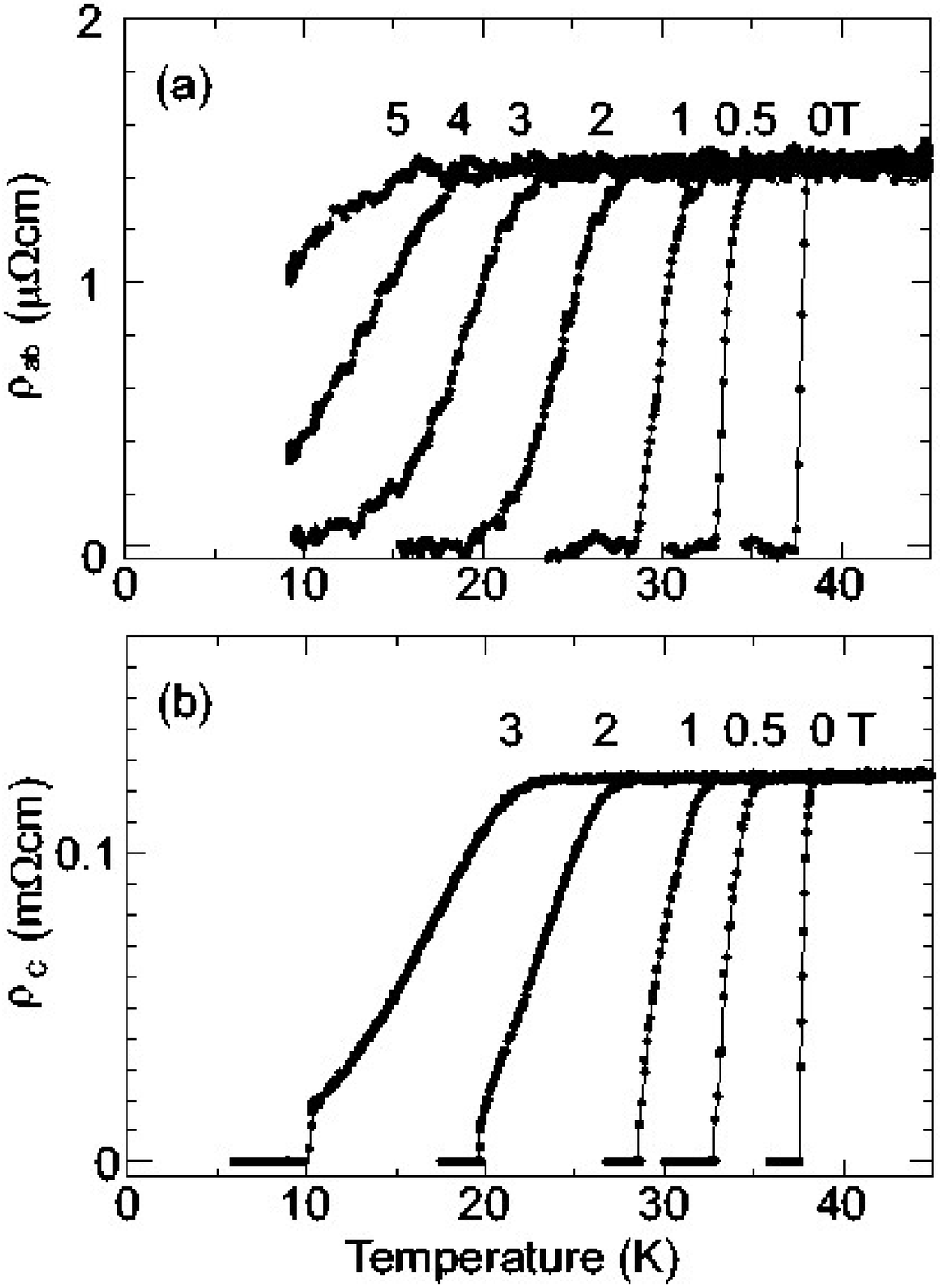}
\end{center}
\caption{\label{fig:rabc_c}
{Resistive broadening in $H//c$ for (a) $I//ab$ and (b) $I//c$. 
In the measurements, $I$=1 mA was used. }}
\end{figure}

\begin{figure}
\begin{center}
\includegraphics[width=10cm]{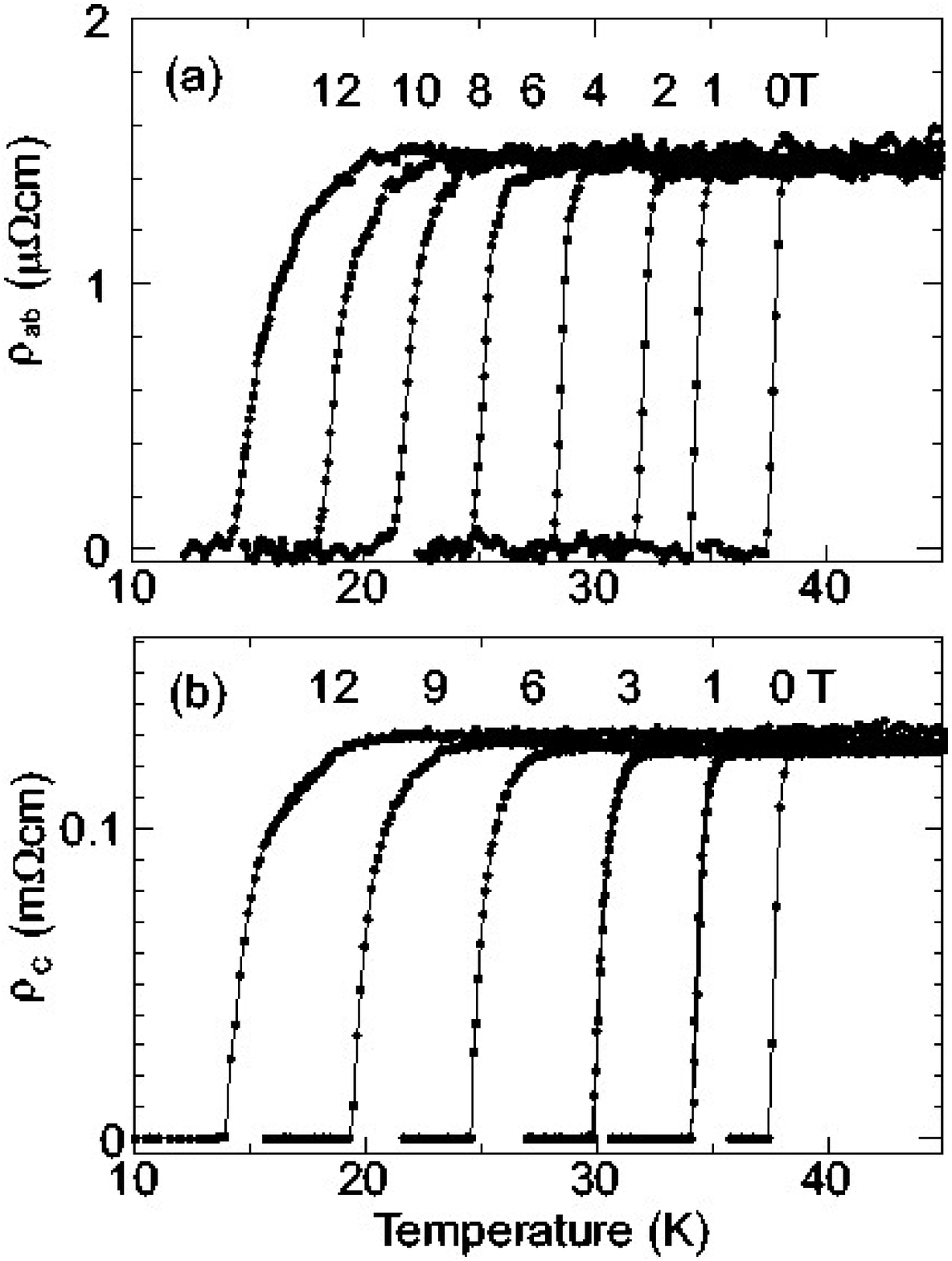}
\end{center}
\caption{\label{fig:rabc_ab}
{Resistive broadening in $H//ab$ for (a) $I//H//ab$ and (b) $I//c$.
In the measurements, $I$=1 mA was used. }}
\end{figure}

Figure \ref{fig:rabc_c}(a) shows the in-plane resistivity ($\rho_{ab}$) 
in magnetic fields parallel to the $c$-axis ($H//c$). 
Although the superconducting transition at $H=0$ is very sharp, 
the transition becomes broader with increasing the magnetic field. 
The resistivity remains finite even at the lowest temperature 
in high magnetic fields. 
The $c$-axis resistivity ($\rho_c$) also shows a broadened SC transition 
in $H//c$ (Fig. \ref{fig:rabc_c}(b)). 
As is similar to the case of $\rho_{ab}$, the resistive broadening in 
$\rho_c$ is enhanced with increasing the magnetic field, and 
the degree of broadening is very similar to that in $\rho_{ab}$. 
On the other hand, the resistive broadening is narrower 
in fields parallel to the $ab$-plane ($H//ab$), 
as is shown in Figs. \ref{fig:rabc_ab} (a) and (b). 
For example, the broadening width $\Delta T_c \approx$ 1 K at $H(//ab)$=2 T, 
while $\Delta T_c \approx$ 7 K at $H(//c)$=2 T.  
Even for $H//ab$, resistive broadening is clearly observed 
at high enough fields. 
It should be noted that the broadening behavior is very similar in 
$\rho_{ab}$ and $\rho_c$. 
The results in Figs. \ref{fig:rabc_c} and \ref{fig:rabc_ab} 
suggest that a broadening width in a magnetic field 
is determined only by the direction of external field, 
being independent of the current direction. 
Therefore, the flux creep by a Lorentz force cannot be the origin 
of resistivity broadening. 
This broadening nature is reminiscent of the vortex properties in 
High-$T_c$ cuprate superconductor \cite{suzuki}. 

\begin{figure}
\begin{center}
\includegraphics[width=10cm]{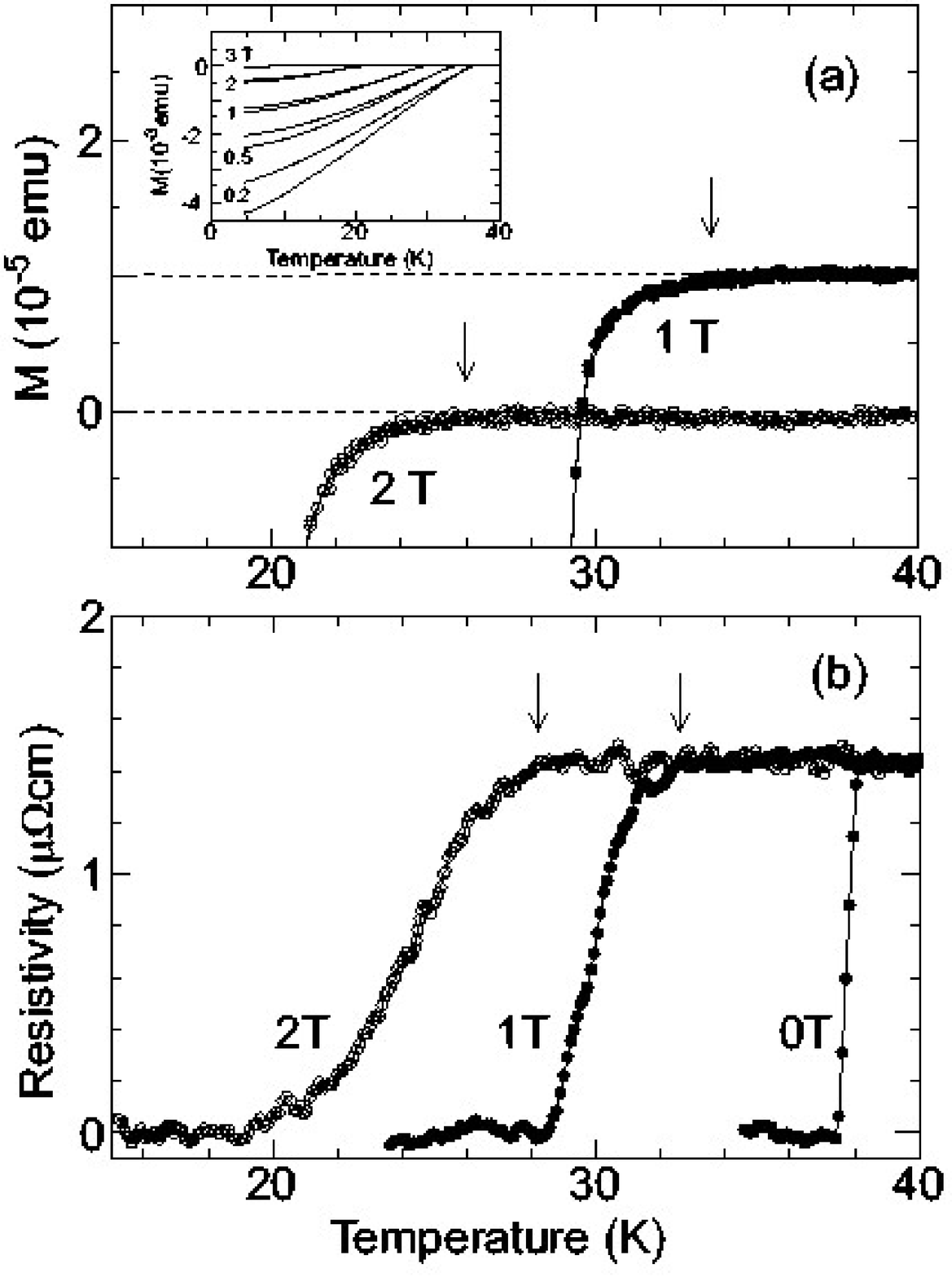}
\end{center}
\caption{\label{fig:hparac}
{(a) Temperature dependence of magnetization in $H//c$ 
around superconductivity transition. 
The data are intentionally shifted upward. 
The inset shows magnetization plotted with a larger scale. 
(b) The resistivity at 1 and 2 T measured with $I$=1 mA. 
Arrows indicate onset temperature of superconductivity transition, 
$T_{on,M}$ or $T_{on,\rho}$.}}
\end{figure}

In order to examine whether the resistivity drop corresponds to 
bulk superconductivity or not, we measured magnetization of the aligned 
mosaic sample. 
Figure \ref{fig:hparac}(a) shows the magnetization for $H//c$ near $T_c$. 
For comparison, the in-plane resistivity ($\rho_{ab}$) is presented 
in Fig. \ref{fig:hparac}(b). 
Broadening of superconductivity transition 
is observed not only in resistivity but also in magnetization. 
The onset temperature of magnetization ($T_{on,M}$) 
is much higher than zero resistivity temperature $T_{0,\rho}$, 
and comparable to $T_{on,\rho}$. 
The diamagnetic signal near $T_{on,M}$ and $T_{on,\rho}$ 
is of the order of 10$^{-6}$ emu for 100 pieces of crystals. 
If only one piece of crystal were measured, the diamagnetic signal could not 
be detected in this $T$-range, 
since the resolution of the magnetometer is of the order of 10$^{-8}$ emu. 
The absolute diamagnetic signal from $T_{on}$ to $T_{0,\rho}$ 
($|M(T_{on,M})-M(T_{0,\rho})|/V$) is of the order of 0.1 Gauss/cm$^3$, 
on the assumption that a whole volume of 
the sample ($V$) contributes to the diamagnetism. 
The diamagnetic signal above $T_{0,\rho}$ was strongly suppressed 
with increasing $H$ and eventually became undetectable at $H \geq$ 4 T, 
although resistivity shows a drop as a sign of superconductivity. 

\begin{figure}
\begin{center}
\includegraphics[width=10cm]{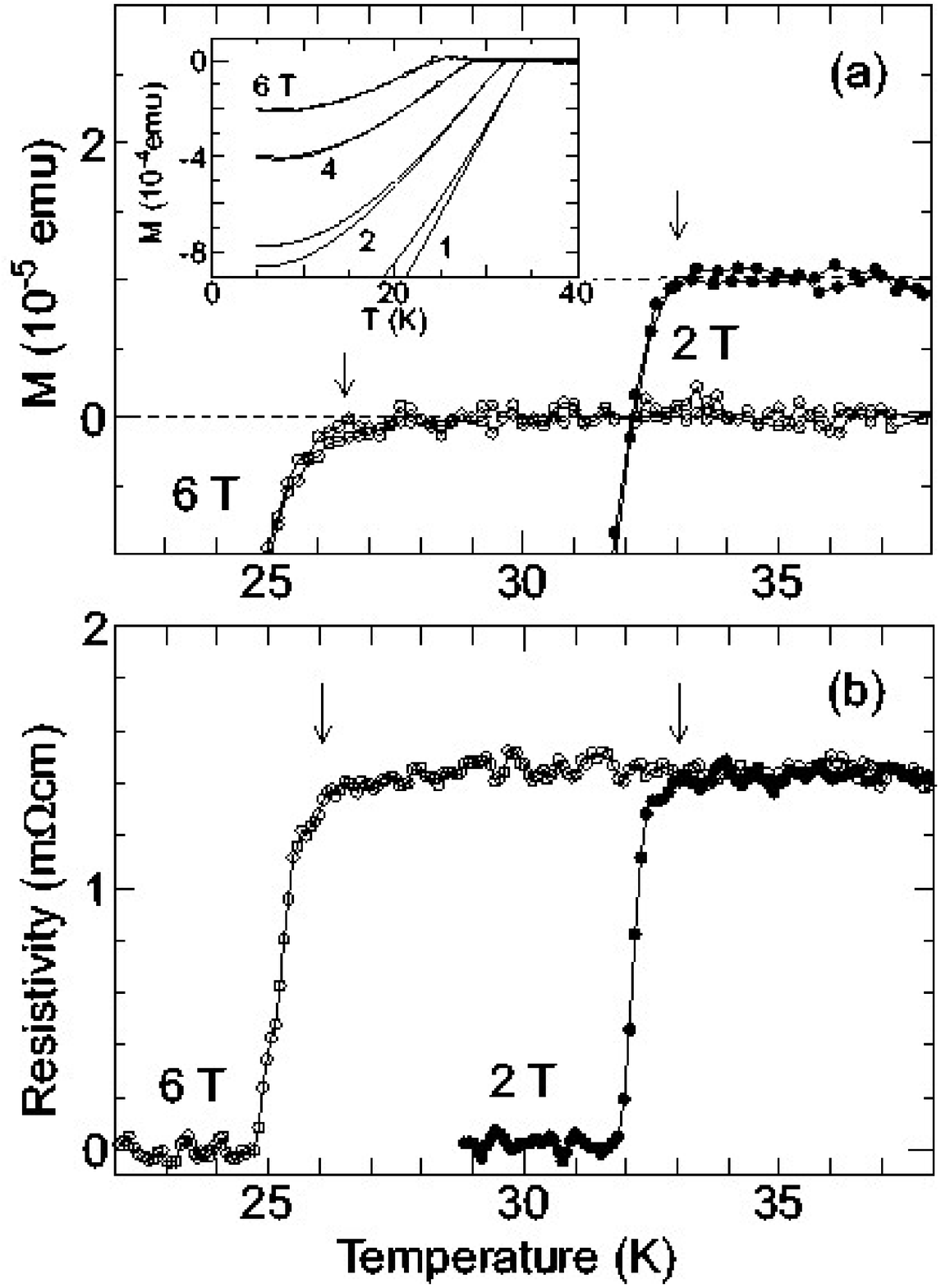}
\end{center}
\caption{\label{fig:hparaab}
{(a) Temperature dependence of magnetization in $H//ab$ 
around superconductivity transition. The data are intentionally 
shifted upward. The inset shows magnetization plotted 
with a larger scale. 
(b) The resistivity at 2 and 6 T, measured with $I$=1 mA. 
Arrows indicate onset temperature of superconductivity transition, 
$T_{on,M}$ or $T_{on,\rho}$.}}
\end{figure}

Also in $H//ab$, the magnetization shows broadening of SC transition, 
but with a narrower width than the case of $H//c$ 
owing to the anisotropy in the superconducting state 
(Fig. \ref{fig:hparaab}). 
In this field direction, $T_{on,M}$ is nearly equal to $T_{on,\rho}$ but 
1$\sim$2 K higher than $T_{0,\rho}$. 
For a polycrystalline MgB$_2$ in which crystal axes of grains 
are randomly oriented, 
the magnetic properties below $T_c$ is dominated by the $H//ab$ component 
because of the higher critical field in this field direction. 
Since a large volume is available for a polycrystalline sample, 
even a small diamagnetic response can easily be detectable. 
This is the reason why a good correspondence of the critical field 
between resistivity and 
magnetization was reported for polycrystalline MgB$_2$ \cite{G_Fuchs}, 
whereas a small signal may have been missed in the measurements 
of single crystal \cite{Welp,Pradhan_02}. 

For discussion, it may be useful to examine the other examples of 
resistive broadening. 
In high-$T_c$ superconducting cuprate (HTSC), 
it is widely accepted that $H_{c2}$ is not well defined, 
owing to a strong thermal fluctuation in comparison to the superconducting 
condensation energy. 
As a consequence of fluctuation, vortex liquid state is observed 
in a wide temperature range. 
For the case of MgB$_2$, 
the coherence length is much shorter than those for 
conventional superconductors. 
It leads to weakness of SC for thermal disturbance. 
The effect of thermal fluctuation on superconductivity is evaluated 
by the condensation energy $E_c$ per coherence volume $V_c$, 
$V_cE_c=\xi_{ab}^2\xi_c N(0) \Delta^2 /2$, 
since nucleation of superconductivity 
occurs fractionally within this volume near $T_c$. 
By using the reported values of $\xi_{ab}$=68\AA, $\xi_c$=23\AA 
\cite{Eltsev_Hc2}, 
$N(0)$=0.25 states/eV-cell \cite{an_pickett}, 
and $2\Delta$=106 cm$^{-1}$ \cite{quilty}, 
we obtain $V_c E_c$ $\sim$ 20 meV. 
This is orders of magnitude lower than those of the conventional 
superconductors. 
Furthermore, if the smaller gap \cite{bouquet,tsuda,quilty_c-axis} 
is taken into account, 
$V_cE_c$ related to this gap becomes the same order as $k_BT_c$=3.4 meV. 
Therefore, it is likely that superconducting fluctuation effect is prominent 
in MgB$_2$, and that the $H_{c2}$ of MgB$_2$ is not well determined. 
Here it should be noted that the specific heat \cite{Welp} and 
the thermal conductivity \cite{Sologbenko2} for single crystal MgB$_2$ 
hardly show a clear superconductivity signal above $T_{0,\rho}$ . 
It suggests that thermodynamical transition temperature 
is defined at $T_{0,\rho}$ rather than at $T_{on,\rho}$. 
In this case, $T_{on}$ is the characteristic temperature 
at which superconducting fluctuation becomes conspicuous. 

The degree of fluctuation can be estimated from the 
criterion $|T_c-T| \leq T_cG_i$, where $G_i=(T_c/H_c^2\xi_c\xi_{ab}^2)/2$ 
is the Ginzberg number and $H_c$ is thermodynamic critical field 
\cite{Blatter}.  
For the present case, the value $T_c G_i$ is much smaller than $10^{-2}$ K, 
suggesting a negligible fluctuation effect at zero magnetic field. 
On the other hand, the situation is different at sufficiently 
high magnetic field $H$, where the cyclotron radius of Cooper pair 
$r_0=(\phi_0/2\pi H)^{0.5}$ becomes shorter 
than the coherence length \cite{tsuneto}. 
The critical temperature region in $H//c$ can be estimated as
$|T_{cH}-T| \leq (k_B H /\Delta C \phi_0 \xi_c)^{2/3} T_{c0}$, where $k_B$ is 
the Boltzmann constant, $\Delta C$ is specific heat jump, 
and $\phi_0$ is a quantum flux. 
Since $\xi_{ab} \sim$ 100 \AA, $r_0$ becomes shorter than $\xi_{ab}$ 
at $H \geq$ 3 T. 
With using $\Delta C/k_B$=3.3$\times 10^{20}$ carriers/cm$^3$ \cite{wang}, 
$H$=3 T, and $\xi_c=23$ {\AA} \cite{Eltsev_Hc2}, 
the critical temperature region of 0.6 K is obtained. 
It is large enough to introduce broadening of 
the superconducting transition in MgB$_2$. 
The narrower broadening in $H//ab$ may also be explained 
by the same model. 
In $H//ab$, $\xi_c$ should be compared with $r_0$. 
Therefore, a larger field $H$ is necessary to manifest fluctuation effect, 
because $\xi_c < \xi_{ab}$. 

\begin{figure}
\begin{center}
\includegraphics[width=10cm]{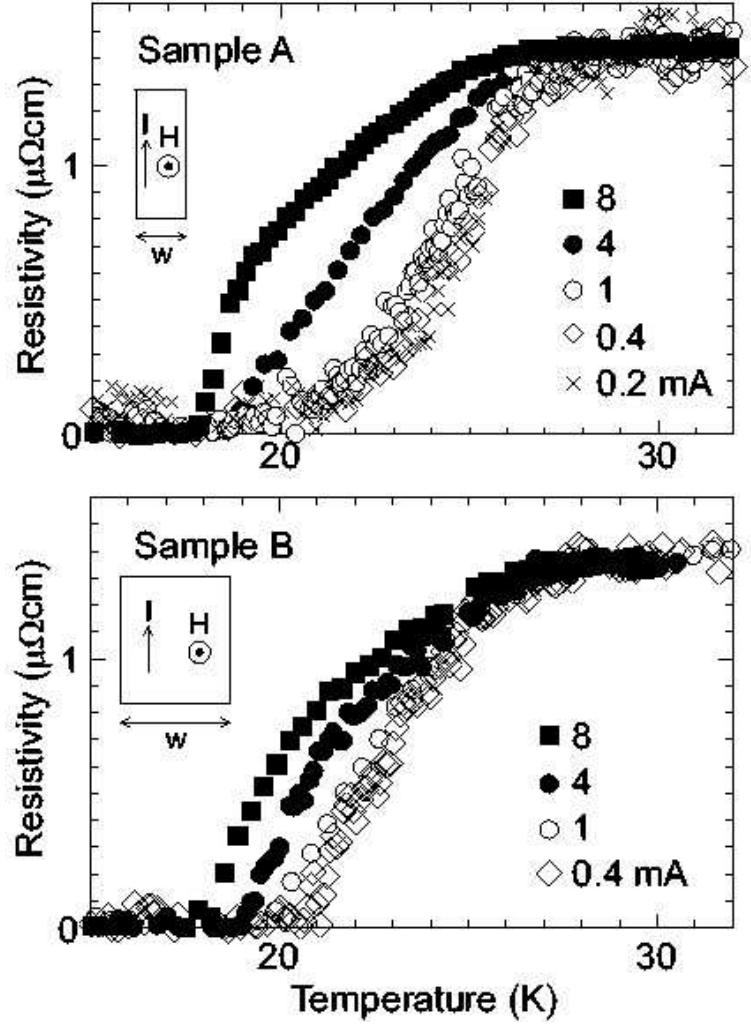}
\end{center}
\caption{\label{fig:idep}{In-plane resistivity ($\rho_{ab}$) in $H$=2 T 
($H//c$), 
measured with different currents for two samples. 
The dimensions for the samples are as follows. 
Sample A: The width $w$=84 $\mu$m, and the thickness along the $c$-axis 
$d$=50 $\mu$m. 
Sample B: $w$=260 $\mu$m, $d$=50 $\mu$m. 
The insets show the schematic view of the sample configuration. 
The current density for $I$=1 mA is $J$=24 A/cm$^2$ in Sample A and 
$J$=7.7 A/cm$^2$ in Sample B, assuming a uniform current flow in the sample. 
}}
\end{figure}

The origin of the resistive broadening has been discussed in relation 
to the non-ohmic behavior \cite{Welp,HJ_Kim,Pradhan_01,Pradhan_02}. 
Figure \ref{fig:idep} shows an example of resistive broadening, 
measured with different current densities. 
In the figure Sample A and Sample B were taken from the same batch, 
and had been preserved in the same dry box. 
The only difference between them is the crystal width. 
Namely the sample dimension perpendicular to the current direction is 
three times larger in Sample B than in Sample A, 
which gives a different current density for the same current value. 
The resistivity curves of these two samples are similar 
when the current $I$=8 mA. 
With decreasing the current, both samples show non-ohmic behavior, 
but more remarkable change of $\rho(T)$-curve is seen in Sample A 
than in Sample B. 
Eventually the $\rho (T)$-curve in sample A shows little change 
with reducing current. 
The resistive curves in the low current regime show a clear difference 
between the samples.
In Sample A, the resistivity steeply decreases as the temperature decreases 
from $T_{on}\approx27$ K, creating a concave curvature, 
while a moderate decrease is seen in Sample B. 
These phenomena strongly suggest 
that the resistive curves are governed by some surface effect, 
since a bulk critical behavior should be determined by the current density. 

A possible explanation for the above current-dependence 
is surface superconductivity \cite{Welp,Danna}. 
However, the observed diamagnetic signal in between $T_{on}$ and $T_{0,\rho}$ 
is too large to be attributed to the surface 
superconductivity \cite{de_Gennes,Abrikosov}. 
Furthermore, the resistive broadening  is observable even in $H//ab$, 
in which surface superconductivity should be negligible. 
Therefore we conclude that the resistivity drop in the critical $T$-region 
originates from a bulk superconducting fluctuation. 

By taking these facts into consideration, we propose a possibility of 
surface barrier effect \cite{D_Fuchs} for the origin of non-ohmic behavior and 
the different resistivity curve with low current 
in Fig. \ref{fig:idep}. 
In this case, a substantial current flows on the side-surface of the crystals 
in a magnetic field because of a high critical current due to 
the surface barrier. 
The temperature dependence of resistivity in $H//c$ 
is determined by the two components, the bulk resistance $R_{bulk}$ 
and surface barrier resistance $R_{sf}$, 
as $1/\rho \sim 1/R_{bulk} + 1/R_{sf}$. 
Since the side surface area is common for the two crystals, 
the difference of the current dependence is ascribed to the $1/R_{bulk}$; 
$1/R_{bulk}$ in Sample A is about three times smaller than that in Sample B 
due to the three times narrower width in the dimensions. 
Correspondingly the effect of $1/R_{sf}$ is more obvious in Sample A. 
Therefore the non-ohmic behavior can be attributed primarily to the change 
in $R_{sf}$ with the current. 
On the other hand, the resistive curve for Sample B, 
in which less contribution of $1/R_{sf}$ is expected, 
shows a similar convex curvature as seen 
in $H//ab$ even with the lowest current, which is reminiscent of 
a sharp decrease of $R_{bulk}$ expected near $T_c$ 
in a mean-field fluctuation. 

\begin{figure}
\begin{center}
\includegraphics[width=10cm]{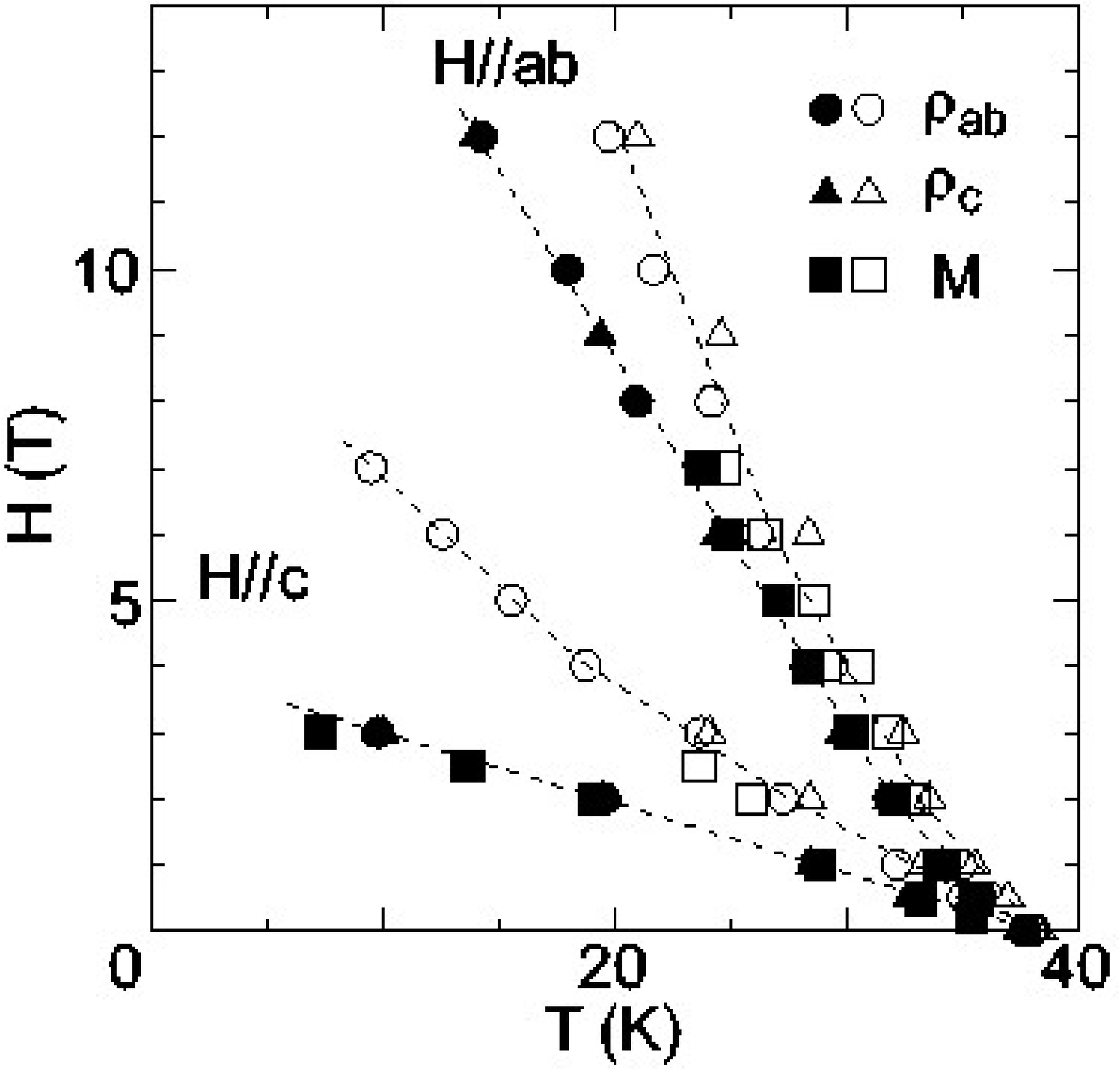}
\end{center}
\caption{\label{fig:ht_phs}{Critical field in $H//ab$ and $H//c$.
Circle, triangle and square represent results obtained from $\rho_{ab}$, 
$\rho_c$, and magnetization $M$, respectively. 
Open and closed symbols represent $T_{on}$ and $T_{0}$, respectively. 
$T_{0}$ for magnetization was determined by the peak position of 
$d^2M/dT^2$. Lines are guides for the eyes. }}
\end{figure}

Finally we show the magnetic phase diagram for MgB$_2$ 
in Fig. \ref{fig:ht_phs}. 
Irrespective of the experimental methods, the data points are 
well on the same curves. 
Here, $T_{0,M}$ is defined by the maximum of $d^2T/dM^2$. 
$T_{on}$ is the characteristic temperature below which superconductivity 
order parameter develops, 
while the critical temperature for a bulk superconductivity 
is close to $T_{0}$. 
Then, the area between $H(T_{on})$ and $H(T_0)$-lines turns out to be 
the superconducting fluctuation region. 
In the magnetization, irreversibility temperatures are not well 
resolved. 

It should be noted that, 
in each direction of magnetic field, the slope of $H_{on}$ is enhanced 
at higher magnetic fields. 
This may be ascribed to the increase of superconducting fluctuation, 
because of the dimensionality change due to the suppression of the 3D-like 
$\pi$-band gap in high magnetic fields. 

\section{Summary}
We report the superconducting transition properties of MgB$_2$ 
in magnetic fields. 
In both resistivity and magnetization, an enhancement of the transition 
broadening was observed with increasing field. 
This suggests that the  broadening is not due to a surface 
superconductivity but to a superconducting fluctuation including 
vortex liquid and/or glass state. 
The effect of superconducting fluctuation is estimated 
from the coherence length and the gap energy.  
We discussed the origin of non-ohmic behavior in $H//c$, 
by comparing the resistivities for the crystals with different 
widths in the dimensions. 
It is most likely that 
the non-ohmic behavior between $T_{on,\rho}$ and $T_{0,\rho}$ is due 
to the surface barrier effect. 

\section{Acknowledgements}
This work was supported by New Energy and 
Industrial Technology Development Organization (NEDO) 
as Collaborative Research and Development of Fundamental Technologies 
for Superconductivity Applications.

\newpage

\newpage

\newpage

\newpage

\newpage

\newpage


\begin{thebibliography}{}

\bibitem{nagamatsu}
J. Nagamatsu, N. Nakagawa, T. Muranaka, Y. Zenitani, J. Akimitsu, 
Nature \textbf{410} (2001) 63.

\bibitem{lee} S. Lee, H. Mori, T. Masui, Y. Eltsev, A. Yamamoto, S. Tajima, J. Phys. Soc. Jpn.  {\bf 70} (2001) 2255.

\bibitem{xu} M. Xu, H. Kitazawa, Y. Takano, J. Ye, K. Nishida, H. Abe, A. Matsushita, G. Kido, Appl. Phys. Lett. {\bf 79} 2779 (2001).

\bibitem{kim} K. Kim, J. Choi, C. Jung, P. Chowdhury, M. Park, H. Kim, J. Kim, Z. Du, E. Choi, M. Kim, W. Kang, S. Lee, G. Sung, J. Lee, 
Phys. Rev. B. {\bf 65} 100510 (R) (2002). 

\bibitem{sologubenko} A.V. Sologubenko, J. Jun, S. M. Kazakov, J. Karpinski, H.R. Ott, Cond-mat/0111273 (unpublished).

\bibitem{Zehetmayer} M. Zehetmayer, M. Eisterner, J. Jun, S. M. Kazakov, J. Karpinski, A. Wisniewski, H. W. Weber, Phys. Rev. B {\bf 66} (2002) 052505. 

\bibitem{Welp} 
U. Welp, G. Karapetrov, W. K. Kwok, G. W. Crabtree, 
Ch. Marcenat, L. Paulius, T. Klein, J. Marcus, 
K. H. P. Kim, C. U. Jung, H.-S. Lee, B. Kang, S.-I. Lee, 
cond-mat/0203337 (unpublished). 

\bibitem{masui_iss} T. Masui, S. Lee, A. Yamamoto, S. Tajima, 
Physica C {\bf 378-381} (2002) 216.

\bibitem{Simon} F. Simon, A. J\'{a}nossy, T. Feh\'{e}r, F. Mur\'{a}nyi, 
S. Garaj, L. Forr\'{o}, C. Petrovic, S. L. Bud'ko, G. Lapertot, 
V. G. Kogan, P. C. Canfield, 
Phys. Rev. Lett. {\bf 87} 047002 (2001). 

\bibitem{HJ_Kim} 
Hyeong-Jin Kim, W. N. Kang, Eun-Mi Choi, Mun-Seog Kim, Kijoon H. P. Kim, 
Sung-Ik Lee, Phys. Rev. Lett. {\bf 87} 087002 (2002). 

\bibitem{Pradhan_01} 
A. K. Pradhan, Z. X. Shi, M. Tokunaga,  T. Tamegai, Y. Takano, K. Togano, H. Kito, H. Ihara, Phys. Rev. B {\bf 64} 212509 (2001). 

\bibitem{Pradhan_02}
A. K. Pradhan, M. Tokunaga, Z. X. Shi,  Y. Takano, K. Togano, H. Kito, 
H. Ihara, T. Tamegai, Phys. Rev. B {\bf 65} 144513 (2002). 

\bibitem{ginsberg} Qiang Li, 
in {\it Physical Properties of High Temperature Superconductors {\bf V}}, 
edited by D. M. Ginsberg (World Scientific, Singapore, 1994), 
pp. 209-264, and references therein. 

\bibitem{ishiguro}
T. Ishiguro, K. Yamaji, G. Saito, {\it Organic Superconductors 
(Second Edition)}
(Springer-Verlag, 1998), and references therin. 

\bibitem{lascialfari} A. Lascialfari, T. Mishonov, A. Rigamonti, P. Tedesco, 
A. Varlamov, Phys. Rev. B. {\bf 65} (2002) 180501(R). 

\bibitem{Park} Tuson Park, M. B. Salamon, C. U. Jung, Min-Seok Park, Kyunghee Kim, Sung-Ik Lee, Cond-mat/0204233 (unpublished). 

\bibitem{suzuki} For example, M. Suzuki, M. Hidaka, Jpn. J. Appl. Phys. 
{\bf 28} L1368 (1989). 

\bibitem{G_Fuchs} G. Fuchs, K. -H. M\"{u}ller, A. Handstein, K. Nenkov, 
V. N. Narozhnyi, D, Eckert, M. Wolf, L. Schutz, 
Solid State Commun, {\bf 118} 497 (2001). 

\bibitem{Eltsev_Hc2} Yu. Eltsev, S. Lee, K. Nakao, N. Chikumoto, S. Tajima, 
N. Koshizuka, M. Murakami, Phys. Rev. B {\bf 65} 140501(R) (2001). 

\bibitem{an_pickett} J. M. An, W. E. Pickett, Phys. Rev. Lett. {\bf 86} (2001) 4366.

\bibitem{quilty}
J.W. Quilty, S. Lee, A. Yamamoto, S. Tajima, Phys. Rev. Lett. {\bf 88} 
087001 (2002). 

\bibitem{bouquet}
F. Bouquet, R. A. Fisher, N. E. Phillips, D. G. Hinks, J. D. Jorgensen, 
Phys Rev. Lett. {\bf 87} 047001 (2001).

\bibitem{tsuda}
S. Tsuda, T. Yokoya, T. Kiss, Y. Takano, K. Togano, H. Kito, H. Ihara, S. Shin, Phys Rev. Lett. {\bf 87} 177006 (2001). 

\bibitem{quilty_c-axis}
J. Quilty, S. Lee, S. Tajima, A. Yamanaka, Cond-mat/0206506 (unpublished). 

\bibitem{Sologbenko2}
A. V. Sologubenko, J. Jun, S. M. Kazakov, J. Karpinski, H. R. Ott, 
Phys. Rev. B {\bf 65} 180505(R) (2002). 

\bibitem{Blatter} For example, G. Blatter, M. V. Feigel'man, 
V. B. Geshkenbein, A. I. Larkin, V. M. Vinokur, Rev. Mod. Phys. 
{\bf 66} 1125 (1994).

\bibitem{tsuneto} T. Tsuneto, {\it Superconductivity and Superfluidity} 
(Cambridge University Press, 1999); R. Ikeda, T. Tsuneto, 
J. Phys. Soc. Jpn. {\bf 58} 1377 (1989). 

\bibitem{wang}
Y. Wang, T. Plackowski, A. Junod, Physica C {\bf 355} 179 (2001).

\bibitem{Danna} G. DfAnna, P. L. Gammel, A. P. Ramirez, U. Yaron, C. S. Oglesby, E. Bucher, D. J. Bishop, Phys. Rev. B {\bf 54} 6583 (1996). 

\bibitem{de_Gennes} P. G. de Gennes, {\it Superconductivity of Metals 
and Alloys} (Benjamin, New York, 1966). 

\bibitem{Abrikosov} A. A. Abrikosov, {\it Fundamentals of the Theory of Metals}, Elsevier Science, (1988), Chapter 18.

\bibitem{D_Fuchs} D. T. Fuchs, R. A. Doyle, E. Zeldov, S. F. W. R. Rycroft, 
T. Tamegai, S. Ooi, M. L. Rappaport, Y. Myasoedov, 
Phys. Rev. Lett. {\bf 81} 3944 (1998), and references therein. 

\end{thebibliography}
\end{document}